\documentclass{PoS}
\usepackage{amsmath,amsfonts}
\def \beq  {\begin{equation}}
\def \eeq  {\end{equation}}
\def \beqar {\begin{eqnarray}}
\def \eeqar {\end{eqnarray}}
\def\sqr#1#2{{\vcenter{\vbox{\hrule height.#2pt
\hbox{\vrule width.#2pt height#1pt \kern#1pt
\vrule width.#2pt}\hrule height.#2pt}}}}

\def\la {{\langle}}
\def\ra {{\rangle}}
\def\vx {{\vec x}}
\def\vy {{\vec y}}

\def\vf {{\varphi}}

\def\bvf {{\bar \varphi}}

\def\Tr {{\rm Tr}}

\def\bA {\bar{A}}

\def\by {\bar{y}}

\def\vx {{\vec x}}
\def\vz {\vec{z}}
\def\vy{\vec{y}}

\def\vu {\vec{u}}
\def\vw {\vec{w}}

\def\del {\partial}
\def\bdel{\bar{\partial}}

\def\e {\epsilon}
\def\d {\delta}

\def\bz {{\bar{z}}}

\def\H {{\cal H}}

\def\G {{\cal G}}

\def\bV{{\bar V}}

\def\vf {{\varphi}}
\def \bvf {{\bar \varphi}}

\def \H {{\cal H}}

\def\half{\textstyle{1\over 2}}

\def \PRL {{Phys. Rev. Lett.}}
\def \PL {{Phys. Lett.}}

\def \NP {{Nucl. Phys.}}

\def \JHEP {{JHEP}}
\def \PR {{Phys. Rev.}}

\title{The Hamiltonian Approach to  Yang-Mills (2+1): An Update and Corrections to String Tension}

\ShortTitle{Yang-Mills (2+1)}

\author{\speaker{V. P. NAIR}\thanks{Supported in part by NSF grant PHY-0855515 and a PSC-CUNY grant}\\
        Physics Department, City College of the CUNY, New York, NY 10031\\
        E-mail: \email{vpn@sci.ccny.cuny.edu}}

\abstract{Yang-Mills theories in 2+1 (or 3) dimensions are interesting as nontrivial gauge theories in their
own right and as effective theories of QCD at high temperatures. I shall review the basics of our
Hamiltonian approach to this theory, emphasizing symmetries with a short update on its status. We
will show that the calculation of the vacuum wave function for Yang-Mills theory in 2+1
dimensions is in the lowest order of a systematic expansion. Expectation values of observables can
be calculated using an effective interacting chiral boson theory, which also leads to a natural
expansion as a double series in the coupling constant (to be interpreted within a resummed perturbation
series) and a particular kinematical factor. The calculation of the first set of corrections in
this expansion shows that the string tension is modified by about $ -0.3\%$ to $-2.8\%$ compared to the
lowest order value. This is in good agreement with lattice estimates.}

\FullConference{International Workshop on QCD Green's Functions, Confinement, and Phenomenology - QCD-TNT09\\
		 September 07 - 11 2009\\
		 ECT Trento, Italy}

\begin{document}

\section{Introduction}

As many of you know, Yang-Mills theory in $2+1$ dimensions has been the focus of research as well as an object of fascination for my collaborators and myself for many years. In the first part of this talk, I shall give an update on the status of our attempts to understand this theory.
In the second part, I shall zero in on our recent calculations of corrections to the string tension. This latter part is more than just a calculation of specific results;
it also formulates a systematic calculational framework which may prove to be a gateway to newer results such as glueball masses.

Let me begin by recalling why this theory is of interest to us, although for a QCD-savvy audience such as this, such a reminder is perhaps redundant. Yang-Mills theories have, of course, been a great puzzle for many decades. Since the realistic case of $3+1$ dimensions is highly nontrivial and difficult, 
physicists have often turned to
the time-honored path of seeking guidance from lower dimensions. Yang-Mills theories
in $1+1$ dimensions are exactly 
solvable, but they have no propagating degrees of freedom and, as a consequence, are a little too trivial for us. In $2+1$ dimensions, there are propagating degrees of freedom, the theory has nontrivial dynamics, yet seems to be within reach of some level of analytical investigation.
This is aided by the fact that the theory has a dimensional coupling constant and, as a related fact, because it is
super-renormalizable.
There is also an interesting physical context in which Yang-Mills theory in $2+1$ dimensions,
in its own right, is useful. One can use Yang-Mills theory in $3$ Euclidean dimensions as the high temperature approximation to real QCD; calculations in the Minkowskian $2+1$ dimensional theory can then be continued to  the Euclidean signature to make predictions about the magnetic screening mass and other related phenomena.

\section{A short review and status report}

Our approach to this problem has been to use Hamiltonian analysis and to solve the Schr\"odinger equation for the vacuum wave function \cite{KKN1, KKN2}. Generally, Hamiltonian methods are not easy to implement in a field theory with its infinite number of degrees of freedom and regularization and renormalization issues. But for this particular problem, it has some advantages, because, at constant time, we are dealing with two-dimensional  gauge fields and, for these, our experience with conformal field theory is a useful guide.

As is standard in a Hamiltonian analysis, we then start with the $A_0 =0$ gauge. The spatial components of the gauge potential can be combined as
$A=A_z = \half ( A_1 +i A_2)$, $\bA = A_\bz = \half (A_1 -i A_2)$. The
crucial first step for us is the parametrization of these potentials as
\beq
A= - \del M ~M^{-1}, \hskip .2in  \bA = M^{\dagger -1} \bdel M^\dagger .
\label{phi4}
\eeq
Here $M$ is an $SL(N, {\mathbb C})$-matrix for an $SU(N)$-gauge theory.
More generally, $M$ is a complex matrix which is an element of $G^{\mathbb C}$, the complexification of $G$
which is the Lie group in which the gauge transformations take values.

Time-independent  gauge
transformations act on 
${M}$ via ,
\beq {M}(\vec{x}) \longrightarrow
g(\vec{x}) {M}(\vec{x})~,  ~~~~         g(\vec{x}) \in SU(N)
\label{sc0}
\eeq
The hermitian matrix $H = M^\dagger M$ is thus gauge-invariant. The fact that the gauge transformations are made homogeneous, as in (\ref{sc0}) is why
the parametrization (\ref{phi4}) is so useful.

Wave functions $\Psi (H)$ are gauge-invariant and depend only on $H = M^\dagger M$; this may be thought of as the statement of the Gauss law.
Further, the Jacobian of the transformation $A, ~ \bA \rightarrow H$ can be explicitly evaluated and the inner product for the wave functions can be obtained as
\beq
\la 1\vert 2\ra
= \int d\mu (H) ~\exp \left( {2c_A S_{wzw}(H)} \right) ~\Psi_1^* \Psi_2
\label{sc1} 
\eeq
where $S_{wzw}(H)$ is the Wess-Zumino-Witten action for
the hermitian field $H$ and $d\mu (H)$ is the Haar measure for $H$
viewed as an element of $SL(N, \mathbb{C}) /SU(N)$. 
The number $c_A$ is the quadratic Casimir value for the adjoint representation;
it is defined in terms of the structure constants $f^{abc}$ (of the Lie algebra) as
$c_A \delta^{ab} = f^{amn} f^{bmn}$ and is equal to $N$ for $SU(N)$.
The WZW action is given by
 \beq
S_{wzw} (H) = {1 \over {2 \pi}} \int \Tr (\partial H \bar{\partial}
H^{-1}) +{i \over {12 \pi}} \int \epsilon ^{\mu \nu \alpha} \Tr (
H^{-1} \partial _{\mu} H~ H^{-1} \partial _{\nu}H~ H^{-1} \partial _{\alpha}H)
\label{sc2}
\eeq
Actually we can strengthen the statements given above a little further and argue that the wave functions $\Psi$ and the Hamiltonian can be taken to be functions of a scaled version of the current of the WZW action, namely, of
\beq
J = {2\over e} \del H~ H^{-1}\label{sc2x}
\eeq
One way to see this is to consider the Wilson loop variable,
\beq
W(C) = \Tr ~{\cal P}  e^{-\oint_C A} = \Tr ~{\cal P} \exp \left({e\over 2} \oint_C J
\right)
\label{sc2y}
\eeq
where, in the second step, we have simply used the parametrization (\ref{phi4}).
Since all gauge-invariant quantities can be constructed from the Wilson loop variable
(by suitable choices of the contour of integration $C$), we may restrict attention to
functions of $J$.

Taking the wave functions to be functions of the current $J$, functional derivatives with respect to $A, \bA$ may be obtained via the chain rule of differentiation.
This leads to the expression for the Hamiltonian operator in terms of the $J$ as
$ {\cal H} =  {\cal H}_0 ~+~{\cal H}_1$ with
 \beqar
{\cal H}_0 &=& m  \int J_a (\vz) {\d \over {\d J_a (\vz)}} + {2\over \pi} \int _{w,z} 
 {1\over (z-w)^2} {\d \over {\d J_a (\vw)}} {\d \over {\d
J_a (\vz)}} + {1\over 2} \int_z : \bdel J^a(z) \bdel J^a(z) :\label{rec1}\\
{\cal H}_1&=&
 i e \int_{w,z} f_{abc} {J^c(w) \over \pi (z-w)} {\d \over {\d J_a (\vw)}} {\d \over {\d
J_b (\vz)}} \nonumber
\eeqar
where $m = e^2c_A/2\pi = e^2N/2\pi $. Notice that the first term of the Hamiltonian
is counting powers of $J$ in the wave function on which it acts, giving an energy
contribution $m$ for each $J$. This will turn out to be the essence of the mass gap in the theory.
\vskip .1in
\hrule
\vskip .1in
\noindent $\underline{An ~aside ~ on ~regularization}$
\vskip .1in
The Hamiltonian operator involves products of $J$'s, $\delta/\delta J$'s and 
nonlocal functions such as $(z-w)^{-1}$. As in any field theory, all calculations have to be done with proper regularization. The one we have used involves a thickening of the Dirac $\delta$-function,
\beq
\delta^{(2)}(u,w) \longrightarrow \sigma (\vu , \vw ; \epsilon ) 
= {1\over \pi \epsilon} \exp \left( - {\vert \vu -\vw\vert^2 \over \epsilon}
\right)
\label{reg1}
\eeq
This leads to the replacement of the unregulated Green's function for $\bdel$ by its regularized version $\bar{\G} _{ma} (\vx,\vy)$; i.e.,
\beqar
\bar{G} (\vec{x},\vec{y}) = {1 \over {\pi (x-y)}} ~\longrightarrow~\bar{\G} _{ma} (\vx,\vy)  &=& \int_u \bar{G} (\vec{x},\vec{u}) \sigma (\vu, \vy; \epsilon )~ H (u, {\bar y})
H^{-1} (y, \by ) \nonumber\\
&\approx&{1\over \pi (x-y)}   \Bigl[ \d _{ma} - e^{-|\vx-\vy|^2/\e} \bigl(
H (x,\by) H^{-1} (y, \by) \bigr) _{ma}\Bigr]
\label{reg2}
\eeqar
The parameter controlling the regularization, $\e$, acts as a short-distance cut-off.
Expression (\ref{reg2}) is to be used on functionals where the point-separation
of various factors is much larger than $\sqrt {\e }$. 

While the parametrization (\ref{phi4}) has many advantages, it has an ambiguity in that
$(M ,~M^{\dagger})$ and $(M \bV (\bz),~V(z) M^{\dagger})$  give the same gauge potentials $A, ~\bA$, where $\bV,~V$ are, respectively, antiholomorphic and holomorphic in the complex coordinates $\bz = x_1 + ix_2$ and $z= x_1 -i x_2$. To avoid this ambiguity of parametrization, physical observables in the theory should have invariance under $H \longrightarrow V(z) H \bV (\bz)$.
The regularization used in (\ref{reg2}) respects this ``holomorphic invariance"; in fact, 
factors like $\bigl( H (x,\by) H^{-1} (y, \by) \bigr)$ appear in (\ref{reg2}) precisely for this reason.

All the results we present here have been checked using regularized expressions, with a single regulator from beginning to end, to avoid possible conflicts between regularizations.
The story of regularization, at the level of calculations, is more involved than 
what we have indicated here; for more details, see \cite{KKN1, KKN2, AKN}.
\vskip .1in
\hrule
\vskip .1in

One interesting result can be read off from what we have done so far.
By comparison with resummed perturbation theory, we can see that
$m$ serves as the propagator mass for gluons (which are generated by $J$). This can then be identified with the magnetic screening mass if we interpret the theory as the high temperature limit of
QCD.
For the gauge group $SU(2)$, $m = e^2/\pi \approx 0.32 e^2$.
There are a number of other calculations for the magnetic screening mass, so it is interesting to compare these values. Table 1 shows that while there is some variation, the values are not outrageously different from $e^2/\pi$.
\begin{table}[!b]
\begin{center}
Table 1. Comparison of magnetic mass calculations
\vskip .05in
\begin{tabular}{|c l |}
\hline\hline
$m/e^2$ & Method\\
\hline
~~~0.25~~~~~& Resummation of perturbation theory \cite{JMC}\\
0.35& Lattice, common factor for glueball masses \cite{owe1}\\
0.51&Lattice, maximal abelian gauge \cite{karsch1}\\
0.52&Lattice, Landau gauge \cite{karsch1}\\
0.44&Lattice, $\lambda_3 =2$ gauge \cite{karsch1}\\
0.38& Resummation of perturbation theory \cite{AN}\\
0.28& Resummation of perturbation theory \cite{BP, JP}\\
0.37&Gauge-invariant lattice definition \cite{owe2}\\
\hline
0.32&Calculation via our Hamiltonian method\\
\hline\hline
\end{tabular}
\end{center}
\end{table}

In our earlier work \cite{KKN1, KKN2}, we also solved the Schr\"odinger equation
${\cal H} \Psi_0 = 0$ for the vacuum wave function $\Psi_0$ to
the leading order in a strong-coupling expansion to obtain
$\Psi_0 = e^{-{\half} S}$, where

\beq
S(H) =  \int  \bdel J^a \left[ { 1 \over {\bigl( m
+ \sqrt{m^2
-\nabla^2 } \bigr)}} \right] \bdel J^a
+{\cal O}(J^3)
\label{sc2a}
\eeq
Notice that if one restricts to modes of $J$ with momentum $\ll e^2$,
\beq
S(H) \approx {1 \over 2 m} \int \bdel J^a \bdel J^a
 = {1\over 4 g^2} \int d^2x~ F^a_{ij} F^a_{ij}\label{sc2b}
\eeq
Thus the computation of expectation values reduces in this limit to a calculation in a Euclidean two-dimensional
Yang-Mills theory with a coupling $g^2 = m e^2$. The expectation value of the Wilson loop variable for the representation $R$ then takes the form
\beq
\la W(C) \ra \sim \exp \left( - \sigma_R {\cal A}_C \right)
\label{wilson}
\eeq
where ${\cal A}_C$ is the area enclosed by the curve $C$. The string tension
$\sigma_R$ is given by
\beq
\sqrt{\sigma_R} ~ = ~e^2 \sqrt{c_A c_R \over4\pi}
\label{tension}
\eeq
This compares favorably, to within $1-3\%$, with the latest lattice calculations \cite{teper1, teper2}, see table 2.
\begin{table}[!b]
\begin{center}
Table 2. Comparison of predictions from (\ref{tension}) with lattice calcuations
\vskip .05in
\begin{tabular}{| p{1.3cm}| p{1.7cm} p{1.5cm} p{1.5cm} p{1.5cm} p{1.5cm} p{1.5cm}| }
\hline\hline
Group&\multicolumn{6}{c|}{Representations}\\
\hline
&k=1& k=2 & k=3 & k=2 &k=3 &k=3\\
&Fund.&antisym&antisym&sym&sym&mixed\\
\hline\hline
$SU(2)$&0.345&&&&&\\
&{0.335}&&&&&\\
\hline
$SU(3)$&0.564&&&&&\\
&{0.553}&&&&&\\
\hline
$SU(4)$&0.772&0.891&&1.196&&\\
&{0.759}&{0.883}&&{1.110}&&\\
\hline
$SU(5)$&0.977&&&&&\\
&{0.966}&&&&&\\
\hline
$SU(6)$&1.180&1.493&1.583&1.784&2.318&1.985\\
&{1.167}&{1.484}&{1.569}&{1.727}&{2.251}&{1.921}\\
\hline
$SU(N)$&0.1995 $N$&&&&&\\
$N\!\!\rightarrow \!\!\infty$&0.1976 $N$&&&&&\\
\hline\hline
\end{tabular}
\vskip .05in
The upper entries follow from (\ref{tension}), the lower entries are the lattice values.
$k$ is the rank of the representation.
\end{center}
\end{table}

I have not described in detail the process of solving the Schr\"odinger
equation; this is because there is another argument which can be used to obtain the results to this order \cite{KKN2}.
For this purpose, absorb the factor $\exp (2 c_A S_{wzw}(H))$
from the inner product into the wave function by $\Psi = e^{-c_A S_{wzw}(H)} \Phi$.
Then the Hamiltonian acting on $\Phi$ is given by
\beq
{\cal H} \longrightarrow {\cal H}_\Phi = e^{c_A S_{wzw}(H)}~{\cal H}~e^{-c_A S_{wzw}(H)}
\label{ham1}
\eeq
We shall now consider this expression when $H$ can be approximated as
$H = e^{t^a \vf^a} \approx 1 + t^a \vf^a +\cdots$; this small $\vf$ limit is appropriate for perturbation theory. The new Hamiltonian is then
\beq
\H_\Phi \approx {1\over 2} \int \left[ -{\delta \over \delta \phi^a \delta \phi^a }
+ \phi^a ( m^2 -\nabla^2 ) \phi^a \right] ~+\cdots
\label{ham2}
\eeq
where $\phi^a = \sqrt{ c_A (-\nabla^2) /8\pi m} ~\vf^a$. 
We see that the leading term in
$\H_\Phi$ corresponds to a free field of mass $m$ 
(actually $dim ~G$ fields, counting the multiplicity
due to the index $a$.) The field $\phi$ may be taken as representing the gluon 
(actually a gauge-invariant version with some resummations involved)
and this is consistent with what we said about the gluon mass
being $m$.

We may now trivially calculate the  vacuum wave function in this approximation as
\beq
\Phi_0 \approx \exp \left[ - {1\over 2} \int \phi^a~ \sqrt{m^2 -\nabla^2} ~\phi^a
\right] \label{ham3}
\eeq
If this is transformed back to the $\Psi$-version, we find
\beq
\Psi_0 \approx  \exp \left[ - {c_A \over \pi m} \int (\bdel \del \vf^a) \left[{1\over m + \sqrt{m^2 - \nabla^2}}\right]
(\bdel \del \vf^a) + \cdots \right]
\label{ham4}
\eeq
This was obtained in an approximation of small $\vf$. However, we do know that the full wave function must be a function of the current $J$. So we can ask the question: What function of $J$ can we write such that it reduces to (\ref{ham4}) in the small $\vf$ approximation when $J \approx (2/e) \del \vf$? The answer, to quadratic order
in $J$, is clearly the solution (\ref{sc2a}).
Notice that this argument only depends on the form of integration measure which, in turn, is determined by a two-dimenional anomaly calculation.
So it is rather robust. Of course, a regularized theory is required so that
we can carry out the transformations $\Psi \leftrightarrow \Phi$ without worrying about additional effects coming up.

Another interesting line of development has been about glueball masses.
The authors of \cite{LMY} have followed the general line of reasoning we have outlined, but used a somewhat different wave function
given by
\beq
\Psi_0 = \exp \left[ - {1\over 4m} \int \bdel J ~K[L]~ \bdel J \right],
\hskip .3in K[L] = {J_2 (4 \sqrt{L})\over \sqrt{L}~ J_1(4\sqrt{L}) }
\label{LMY}
\eeq
where $L = {\cal D}\bdel/m^2$ and $J_1, ~J_2$ are Bessel functions of orders
$1$ and $2$ respectively.
The kernel $K$ is, despite appearances, very close to the kernel in (\ref{sc2a}), as is clear from the graphical comparison in figure 1 (I thank A. Yelnikov for this picture). Glueball masses were obtained by calculating the two-point function for different color-singlet composite operators,
characterized by spin, parity and charge conjugation properties ($j^{PC}$-notation). The results, in units of $\sqrt{\sigma_F}$, are shown in table 3. Again, there is reasonable agreement with the lattice data
of reference \cite{teper1, meyer}.
\begin{figure}[!t]
\begin{center}
\includegraphics[height = .35\textwidth, width=.5\textwidth]{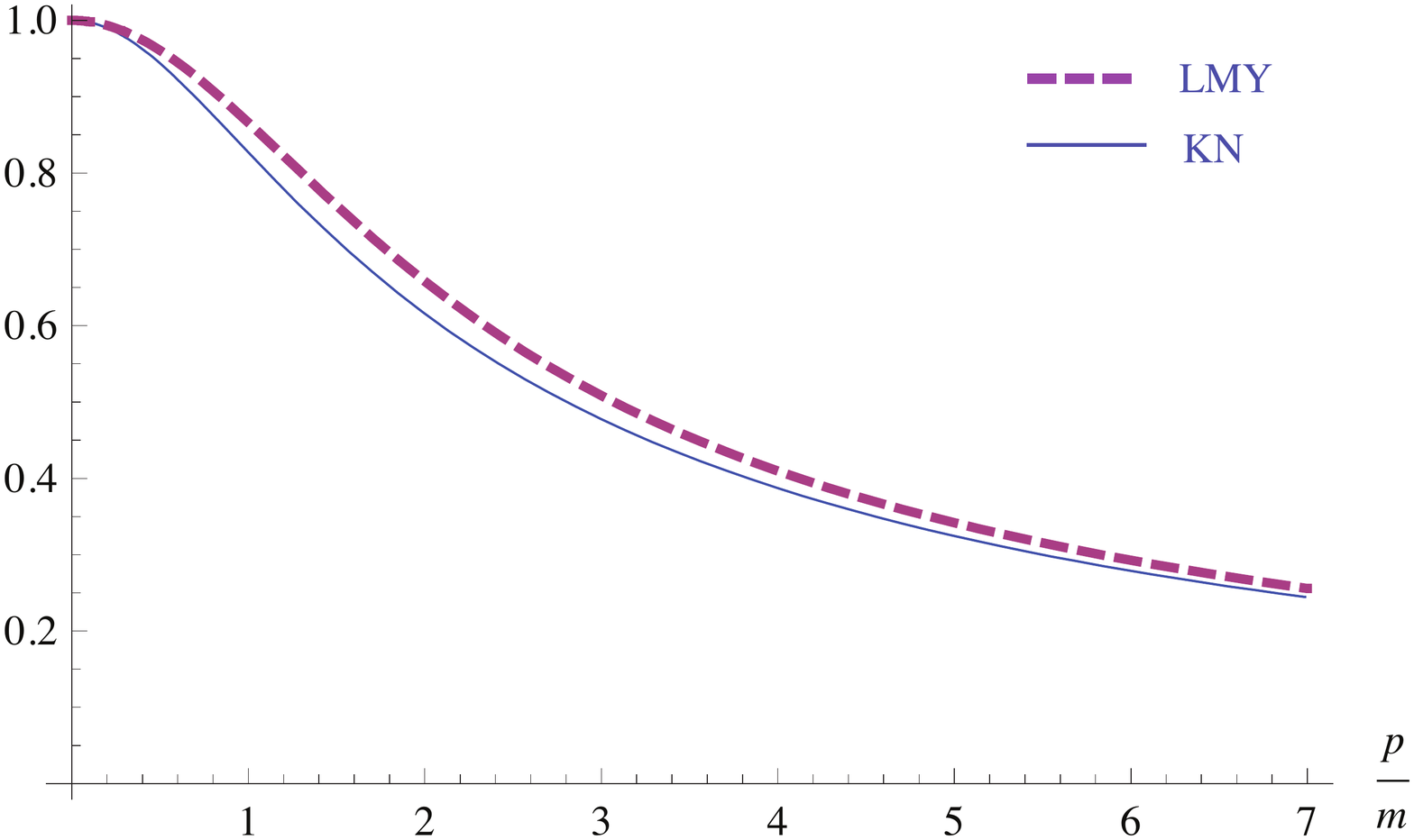}\\
Figure 1. Comparison of KKN and LMY kernels for the Gaussian term of the vacuum wave function as a function of the momentum $p$
\vspace{-24pt}
\end{center}
\end{figure}

\begin{table}[!b]
\begin{center}
Table 3. Comparison of glueball mass estimates from (\ref{LMY}) and lattice calculations.
\vskip .05in
\begin{tabular}{| p{2cm} c p{4cm}  |}
\hline\hline
State&~~~~~LMY Calculation~~~~~~~& Lattice\\
\hline
$0^{++}$&$4.098$& $ 4.065\pm 0.055$\\
$0^{++*}$&$5.407$&$ 6.18\pm 0.13$\\
$0^{++**}$&$6.716$&$ 7.99\pm 0.22$\\
$0^{++***}$&$7.994$&$9.44 \pm 0.38$\\
\hline
$0^{--}$&$6.15$&$ 5.91\pm 0.25$\\
$0^{--*}$&$7.46$&$7.63\pm 0.37$\\
$0^{--**}$&$8.77$&$ 8.96\pm 0.65$\\
\hline
$2^{++}$&$6.72$&$ 6.88\pm 0.16$\\
$2^{++*}$&$7.99$&$ 8.62\pm 0.38$\\
$2^{++**}$&$9.26$&$ 9.22\pm 0.32$\\
\hline
$2^{+-}$&$8.76$&$ 8.04\pm 0.50$\\
$2^{--}$&$8.76$&$ 7.89\pm 0.35$\\
$2^{+-*}$&$10.04$&$ 9.97\pm 0.91$\\
$2^{--*}$&$10.04$&$9.46\pm0.46$\\
\hline\hline
\end{tabular}
\end{center}
\end{table}

There have been a number of other developments as well. The formalism has been extended to the Yang-Mills-Chern-Simons theory and issues related to dynamical mass generation and screening of Wilson loops have been analyzed
\cite{KKN3}. 

Another interesting question has been about the screening of ${\mathbb Z}_N$-invariant representations. The energy of a glue-lump state made of a heavy scalar field and the gluons can be estimated in the limit
of large $e^2$ and large $N$ \cite{AKN}. This gives the energy at which an adjoint string breaks as $E_* \approx 7.92 ~m$, for $SU(2)$, to be compared with the lattice value of $E_* \approx 8.68 ~m$ \cite{For}; we are off by about $8.8\%$. Considering the difficulties in calculating both in our method and the lattice, this may be taken as not too bad. The calculation is more a qualitative
demonstration of feasibility: it is possible to incorporate screening in our formalism.

We have also worked out the extension of the formalism to the case of space being $S^2$, i.e., for ${\mathbb R}\times S^2$ \cite{AN2}, as a prelude to
calculations on the torus which may be used to analyze finite-temperature effects.
In this regard, I also mention the recent paper by Abe, where some of the torus formalism is developed and  arguments on deconfinement are presented \cite{Abe}.

\section{Corrections to string tension}

In the second half of the talk, I shall focus on corrections to the string tension formula (\ref{tension}) \cite{KNY}. 
This is important for several reasons. First, we want to see the vacuum wave function (\ref{sc2a}) and the corresponding string tension (\ref{tension}) as
the lowest order in a systematic expansion scheme. Secondly, notice that, in the large $N$ limit, our formula differs from the lattice value by about $1\%$.
A different lattice calculation shows the difference to be about $1.55\%$ \cite{kiskis}.
In both cases, the lattice calculations are considered accurate enough that the deviations are statistically significant. So we need a systematic method which has the potential to calculate corrections.

There are two types of corrections possible as exemplified by the diagrams
in figure 2. For the first set, we can calculate the corrections to the propagators (or the coefficient of $\bdel J ~\bdel J$ in the wave function) and then use the result in the calculation of the Wilson loop. This set of corrections will be independent of the representation $R$ of the Wilson line.
The second set would involve vertex type corrections and can be sensitive to the representation of the Wilson line. We will leave out these for now, focusing on the first set. (But note that the second type are related to screening and string breaking problems, which we have commented on already. Also, since these $R$-dependent corrections are left out, comparison with lattice analysis is best made at large $N$.)
\begin{figure}[!b]
\begin{center}
\includegraphics[height = .2\textwidth, width=.45\textwidth]{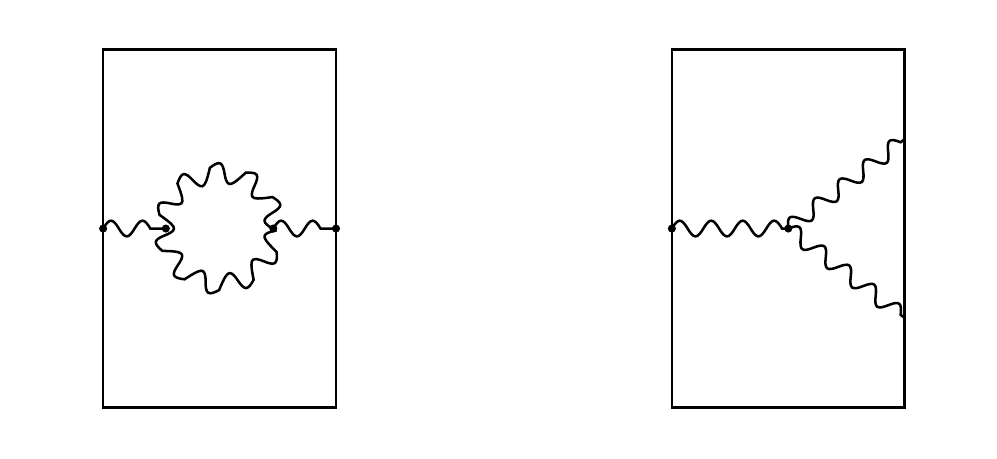}\\
Figure 2. Corrections to Wilson loop (rectangle); wavy lines are
$\la J J\ra$ propagators. 
\end{center}
\end{figure}

The corrections are expected to be purely numerical, so, at first glance, seem to be impossible to control in any way. However, we shall see that a systematic expansion is possible along the following lines.\\
$\underline{Step~ 1}$\\
We rewrite the derivation of the vacuum wave function as a recursive procedure for the solution of the Schr\"odinger equation from which it will be clear that (\ref{sc2a}) is the lowest order result
in a systematic expansion.
For this purpose, we will treat $m$ and $e$ as independent
parameters, setting $m = e^2 c_A/2\pi$ only at the end of all calculations.
For $\Psi_0^* \Psi_0 = e^F$, this gives $F$ as
a power series in $e$, of the form
\beqar
F&=& \int f^{(2)}_{a_1 a_2}(x_1, x_2)\ J^{a_1}(x_1) J^{a_2}(x_2) ~+~
\frac{e}{2}\ f^{(3)}_{a_1 a_2 a_3}(x_1, x_2, x_3)\ J^{a_1}(x_1) J^{a_2}(x_2) J^{a_3}(x_3)
\nonumber\\
&&\hskip .2in~+~
\frac{e^2}{4}\ f^{(4)}_{a_1 a_2 a_3 a_4}(x_1, x_2, x_3, x_4)\ J^{a_1}(x_1) J^{a_2}(x_2) J^{a_3}(x_3)
J^{a_4}(x_4)~+~\cdots\label{exp1}
\eeqar
$f^{(2)}$, $f^{(3)}$, etc., are determined recursively in powers of
$e$, with each function having an expansion
\beqar
f^{(2)}_{\ a_1 a_2}(x_1, x_2)  &=& f^{(2)}_{0\ a_1 a_2}(x_1, x_2)   ~+~ e^2~ 
f^{(2)}_{2\ a_1 a_2}(x_1, x_2)   ~+\cdots\nonumber\\
f^{(3)}_{0\ a_1 a_2 a_3}(x_1, x_2, x_3)  &=& f^{(3)}_{0\ a_1 a_2 a_3}(x_1, x_2, x_3)  ~+~e^2~
f^{(3)}_{2\ a_1 a_2 a_3}(x_1, x_2, x_3) ~+\cdots, ~~{\rm etc.}
\label{exp1a}
\eeqar
To the lowest order, with $q,{\bar q}$ denoting the momenta,
\beq
f^{(2)}_{0\ a_1 a_2}(x_1, x_2)  = \delta_{a_1 a_2}
\left[- {{\bar q}^2 / (m +E_q)}\right]_{x_1, x_2} \approx \delta_{a_1 a_2}~ {{\bar q}^2 /2m} \label{exp2}
\eeq
which leads to (\ref{sc2a}). To calculate corrections to order $e^2$, we need the lowest order results for $f^{(3)}$ and $f^{(4)}$. For example,
\beq
f^{(3)}_{0\ a_1 a_2 a_3}(k_1, k_2, k_3) = -\frac{f^{a_1 a_2 a_3}}{24}\ (2\pi)^2 \delta (\sum k_i)~
 \frac{16}{(\sum_i E_{k_i})}\left[ \frac{\bar k_1 \bar k_2 (\bar k_1 - \bar k_2)}{(m+E_{k_1})(m+E_{k_2})} + {cycl.\ perm.}\right]\label{exp3}
\eeq
There is a similar, but more involved, expression for $f^{(4)}$, see \cite{KNY}.\\
$\underline{Step ~2}$\\
We are interested in corrections to $f^{(2)}$ since that determines the string tension.
From the recursive solution of the Schr\"odinger equation, we find
\beq
e^2~f_2^{(2)}(q) = \frac{m}{E_q}\int \frac{d^2 k}{32\pi}\ \left(\frac{1}{\bar k}\ g^{(3)}(q,k,-k-q)\ + \frac{k}{2 \bar k}\ g^{(4)}(q,k;-q,-k)\right)
\approx  \frac{\bar q^2}{2 m}\,(1.1308) +\ldots\label{exp4}
\eeq
Seemingly, this is a $113\%$ correction, but, as we shall see, other "loop" corrections are important.

We want to emphasize that, even though we use recursion in $e$, our procedure
is different from perturbation theory since $m$ is included exactly in the lowest order result for $F$; more on this later.\\
$\underline{Step ~3}$\\
Since the measure of integration has the WZW action, we can transform the functional integration over $\Psi_0^* \Psi_0 = e^F$ into the integration over a two-dimensional chiral boson field $\vf, ~{\bar \vf}$ (not the $\vf$ we used in parametrizing $H$ as 
$e^{t^a\vf^a}$). In other words, for an observable ${\cal O}$,
\beq
\la {\cal O} \ra = \int {d\mu (H)~ e^{2c_A S_{wzw}(H) }}~e^{F(J)} ~{\cal O}(J)
= \int [d\vf d{\bar \vf}]~  e^{-S(\vf )}~{\cal O}(\sqrt{2\pi /mc_A} ~\bvf t^a \vf )\label{exp5}
\eeq
where the action $S(\vf)$ is given by
\beq
S(\vf ) = \int ~(Z_2 \bvf \bdel \vf + Z_1 \bvf {\bar C}\vf )~-~ F(Z_1\sqrt{2\pi /mc_A} ~\bvf t^a \vf )\label{exp6}
\eeq
($Z_1, Z_2$ are renormalization constants, we will not discuss these further here; see
\cite{KNY}.)\footnote{${\bar C}$ may be set to zero at this point, it was included to show how $Z_1$, $Z_2$ appear in $S (\vf )$.
Also we may note that the transformation to $\vf, ~{\bar\vf}$ is analogous to the fermionization of the WZW model.} Notice that, effectively, the current
$J^a$ is replaced by $Z_1\sqrt{2\pi /mc_A} ~\bvf t^a \vf$.
The function $F(Z_1\sqrt{2\pi /mc_A} ~\bvf t^a \vf )$ contains vertices, $F^{(2)}$ with two currents
(quartic in $\vf$, ${\bar\vf}$), $F^{(3)}$ with three currents, etc.
 For example, we may diagrammatically represent $F^{(2)}$, with two $\vf$'s and two ${\bar\vf}$'s, as
\begin{center}
\begin{tabular}{c c}
{\begin{tabular}{c}
$F^{(2)}= {2\pi \over mc_A}\int ({\bar\vf}t^a\vf)_x ~f^{(2)} (x, y)({\bar\vf}t^a\vf)_y ~=$ \\
{~}\\
\end{tabular} }&
\includegraphics[height = .06\textwidth, width=.13\textwidth]{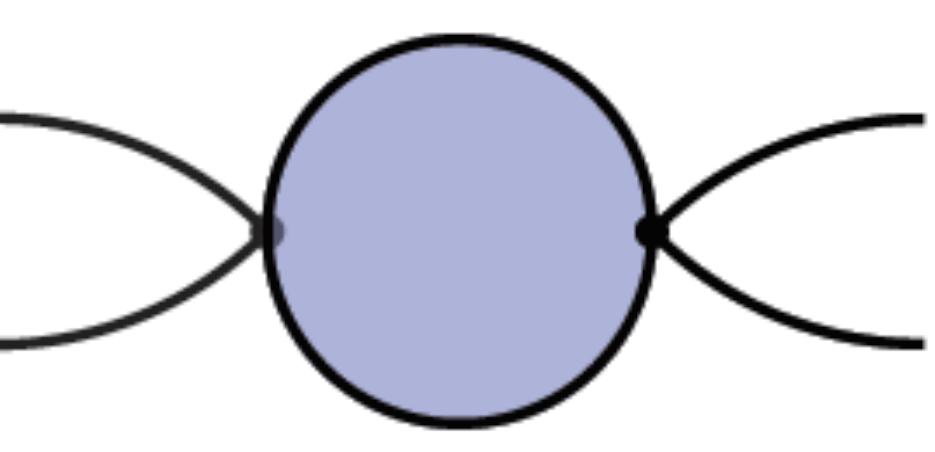}\\
\end{tabular}
\end{center}
The corrections to $F^{(2)}$, which is what we are interested in, may be viewed as Wilsonian renormalization corrections to the quartic vertex in this two-dimensional field theory.\\
$\underline{Step ~4}$\\
In computing the corrections to $F^{(2)}$, we can treat the vertices
$F^{(3)}$, $F^{(4)}$, etc., perturbatively since they carry powers of $e$. However, the lowest term in the vertex $F^{(2)}$, corresponding to
$f^{(2)}_{0a_1a_2}(x_1, x_2)$, has no powers of $e$ and hence its contributions must be summed up. This means that the current-current correlator becomes
\beq
\la \bvf t^a \vf (x)~ \bvf t^b \vf (y)\ra = 
\delta^{ab} {c_A \over \pi} \int {d^2k \over (2\pi )^2} e^{ik(x-y)}~
 {k\over {\bar k}}~ {\left( {m\over E_k}\right)}\label{exp7}
 \eeq
Here $E_k = \sqrt{k^2 +m^2}$; the $m/E_k$ factor arises from the summation of corrections due to $F^{(2)}_0$, shown diagrammatically in figure 3.
\begin{figure}[!t]
\begin{center}
\includegraphics[height = .15\textwidth, width=.7\textwidth]{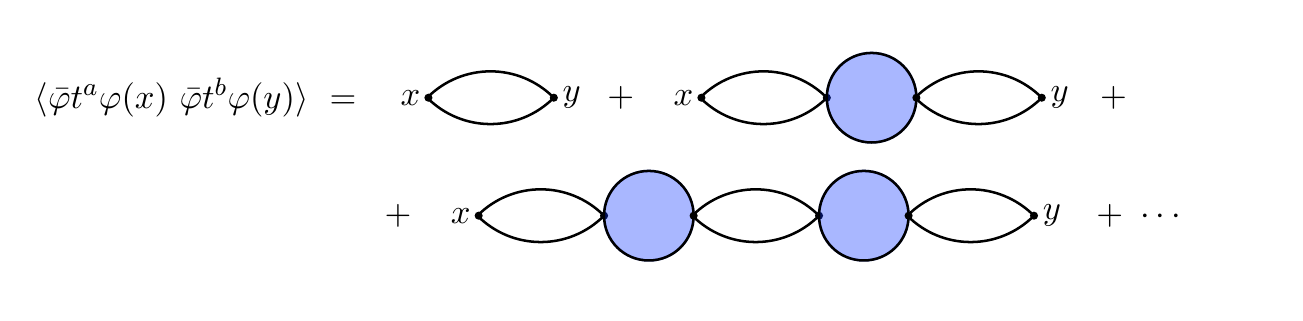}\\
Figure 3. The current-current correlator including all contributions from $F^{(2)}$
\end{center}
\end{figure}
Our computational strategy may then be summarized as follows:
\begin{enumerate}
\item Construct loop diagrams  generated by $F^{(3)}$ (3 factors of ${\bar\vf}t^a \vf$) and $F^{(4)}$
(4 factors of ${\bar\vf}t^a \vf$).
\item They can have arbitrary insertions of $F^{(2)}$'s, leading to a factor of $m/E_k$, as in figure 4.
\item There are also renormalizations (due to $F^{(2)}$) we have to take into account.
\item Sum up $F^{(2)}$ insertions in all diagrams (of order $e^2$) generated by $F^{(3)}$
and $F^{(4)}$.
\item Classify and group these by the number of factors of $m/E_k$.
\end{enumerate}
\begin{figure}[!t]
\begin{center}
\includegraphics[height = .13\textwidth, width=.8\textwidth]{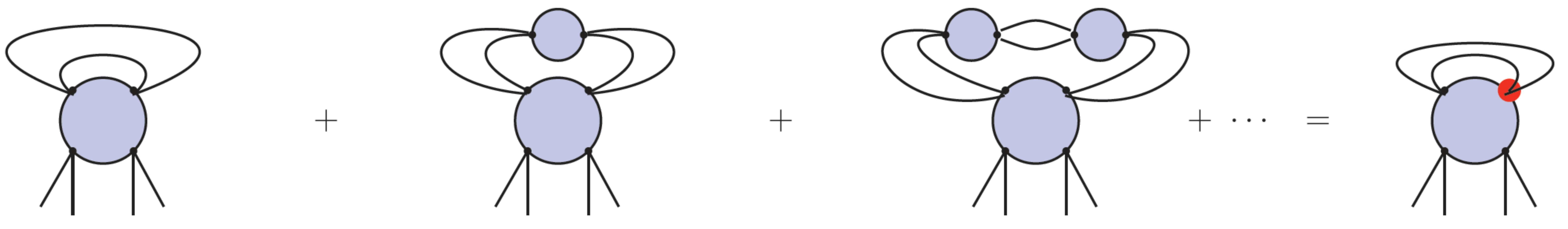}\\
\includegraphics[height = .12\textwidth, width=.35\textwidth]{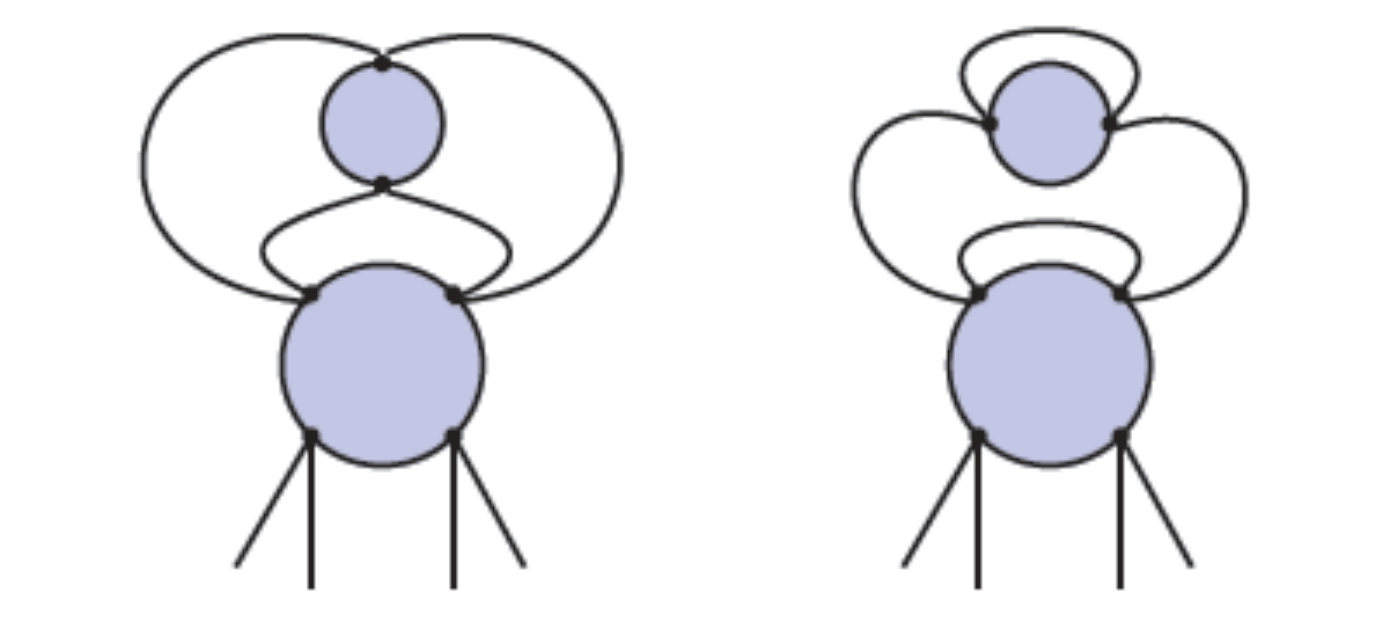} \\
Figure 4. Corrections from $F^{(2)}$ summed up as a factor of $m/E_k$ (shaded
circle
at vertex) and sample renormalization diagrams
\end{center}
\end{figure}

We will compute corrections to {order $e^2$} and up to $4$ powers of $m/E_k$.
Denoting the factors of $m/E_k$ by shaded circles at the vertices, the corrections
to the low momentum limit of $f^{(2)}$ may summarized as in figure 5.
We show the coefficients of ${\bar q}^2/2m$, for small $q, ~{\bar q}$.
\begin{table}[!t]
\begin{center}
\begin{tabular}{c c c c}
\scalebox{.78}{\includegraphics[height = .13\textwidth, width=.13\textwidth]{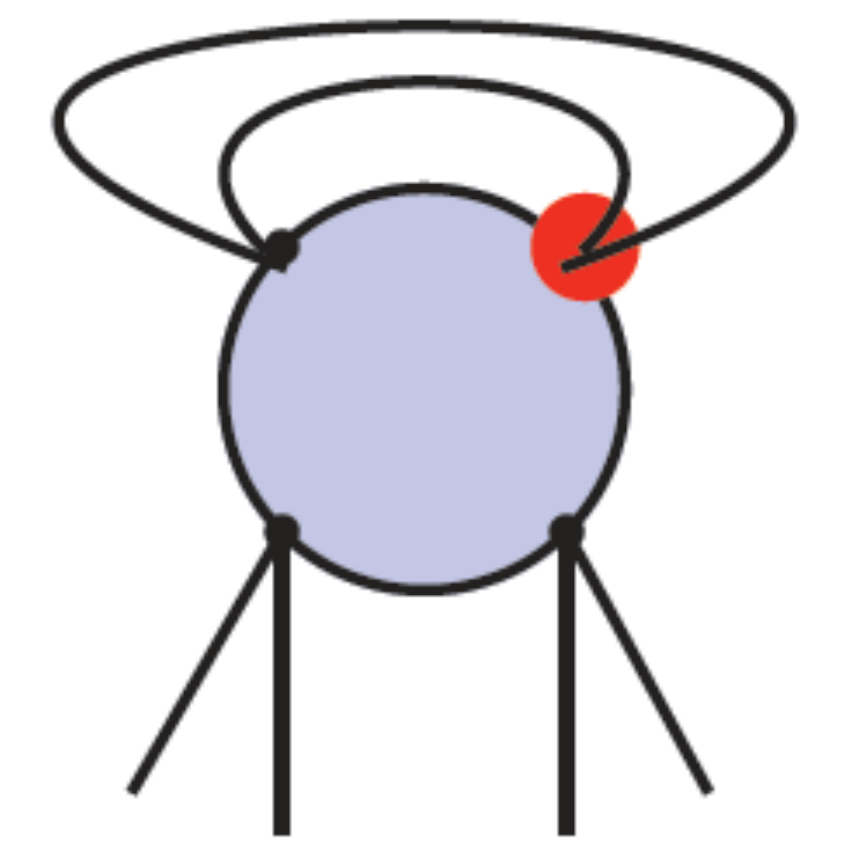}}&\scalebox{.85}{\includegraphics[height = .13\textwidth, width=.13\textwidth]{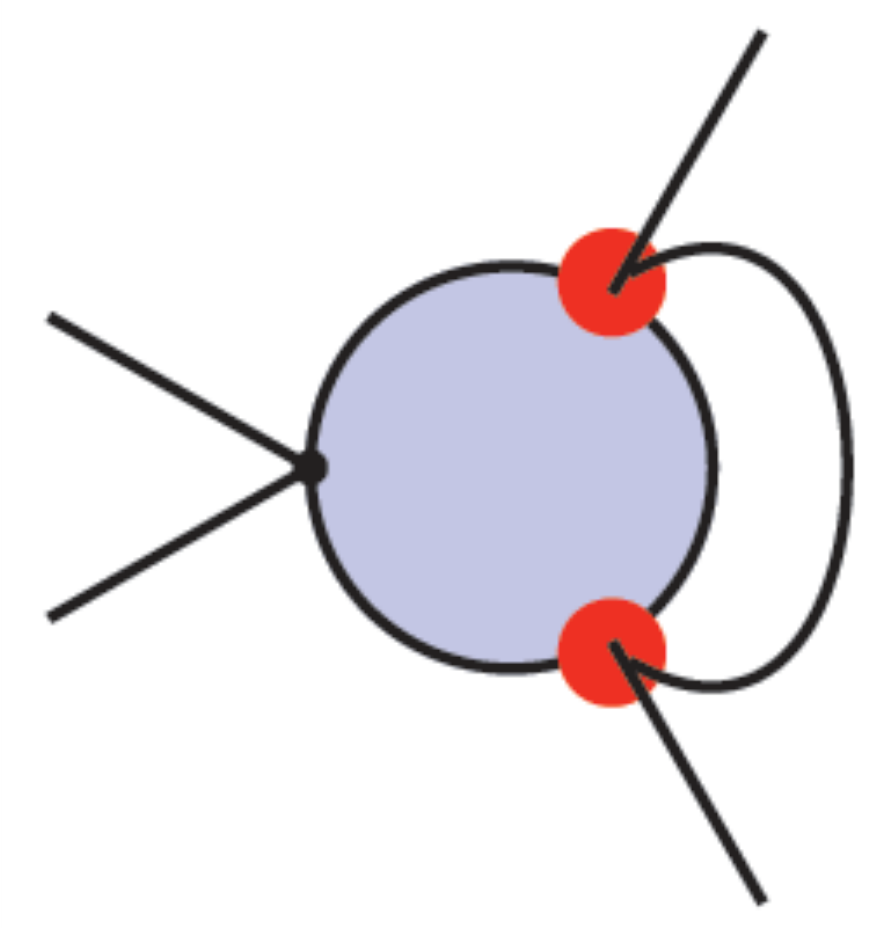}}&
\scalebox{.75}{\includegraphics[height = .11\textwidth, width=.24\textwidth]{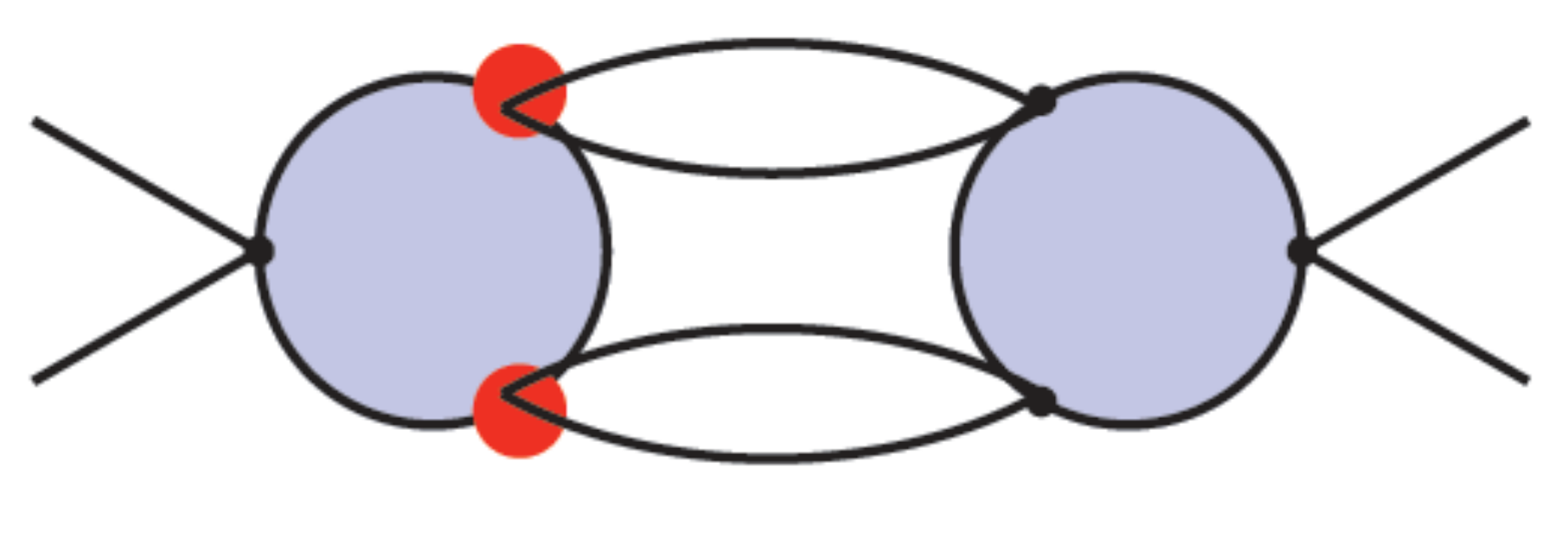}}&\scalebox{.75}{\includegraphics[height = .13\textwidth, width=.13\textwidth]{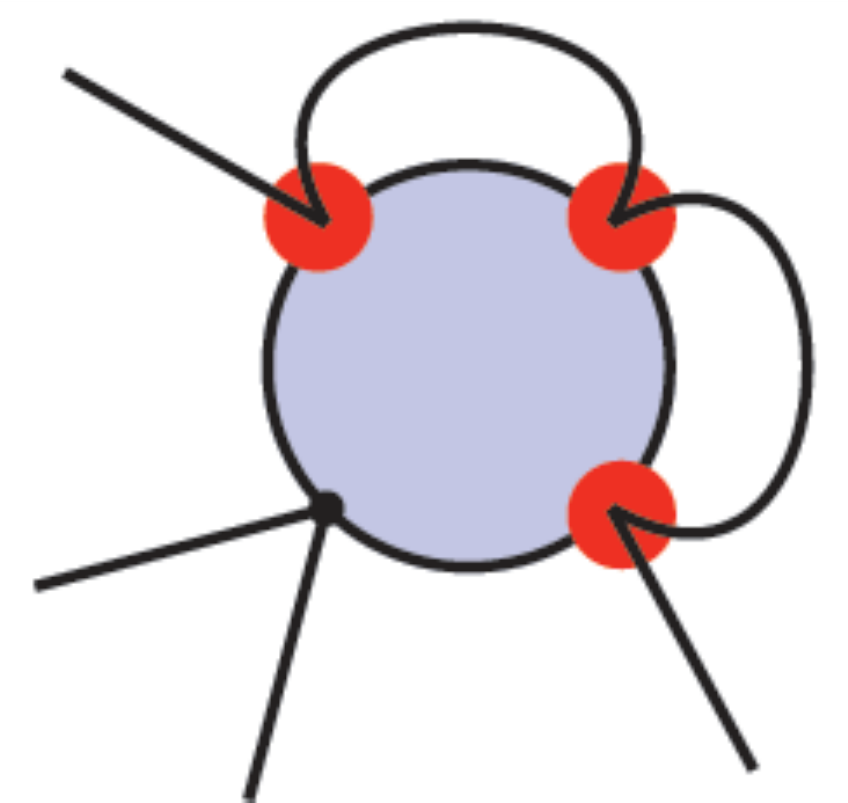}}\\
~~~~~~~$-0.58118 $~~~~~~~&
~~~~~~~$-0.47835$~~~~~~~&
~~~~~~~$0.20169$~~~~~~~&~~~~~~~$-0.23569$~~~~~~~\\
\end{tabular}\\
\vskip .1in
\begin{tabular}{c c c}
\scalebox{1.1}{\includegraphics[height = .06\textwidth, width=.18\textwidth]{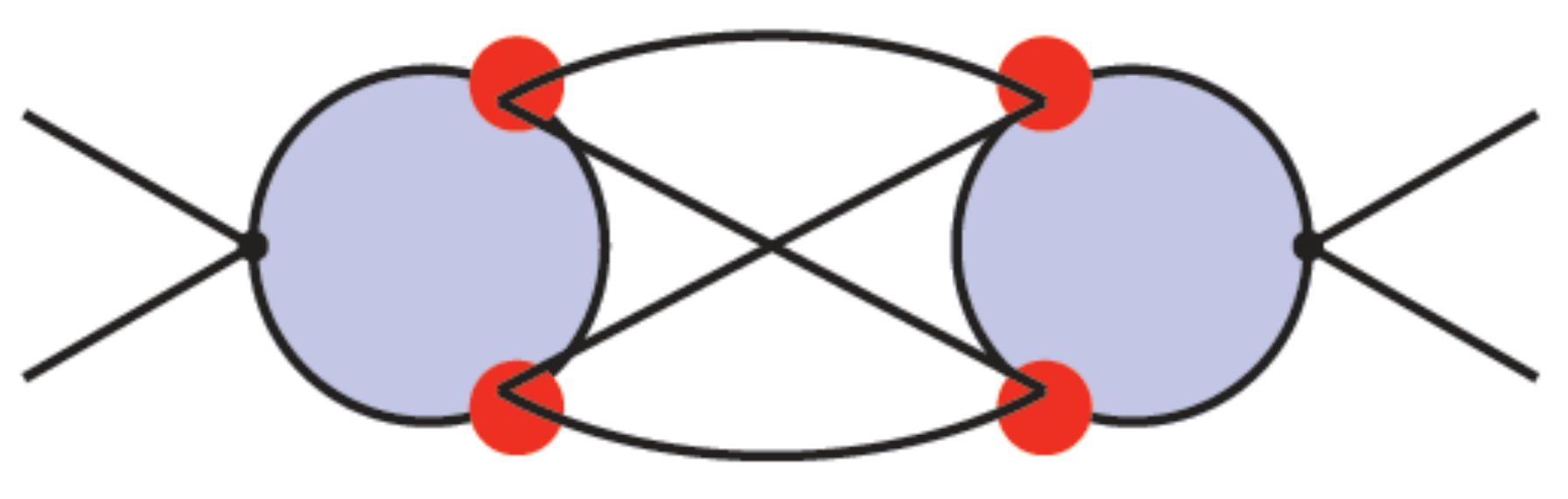}}~~~~~&
\scalebox{1.1}{\includegraphics[height = .07\textwidth, width=.2\textwidth]{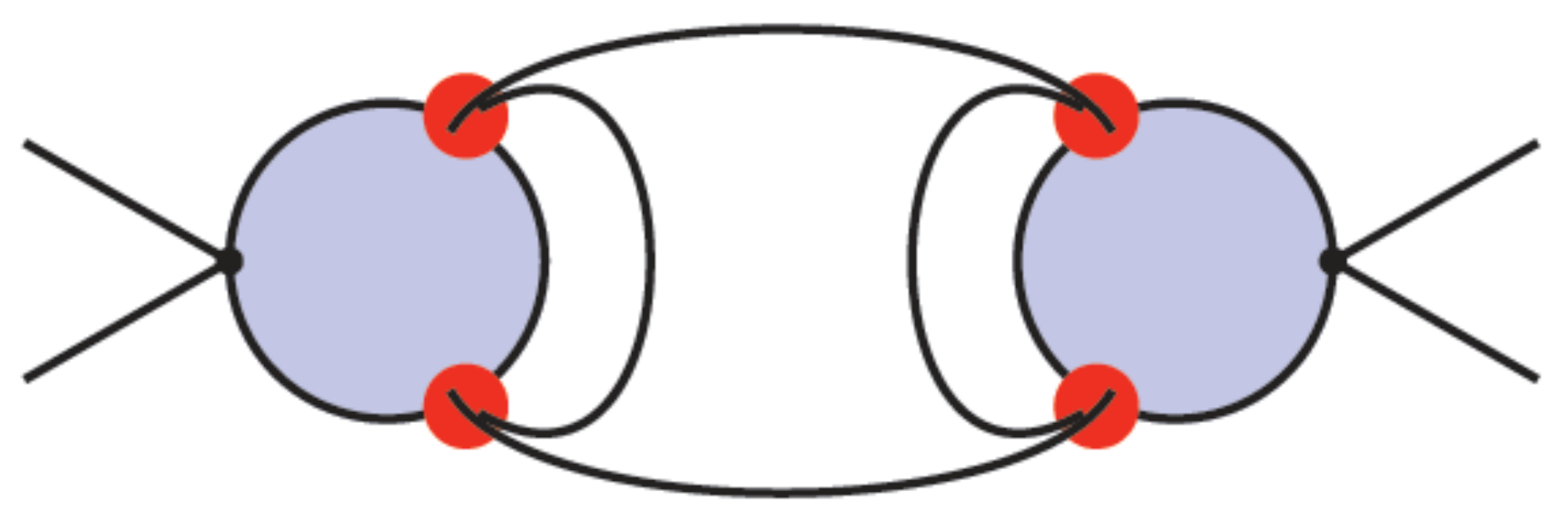}}~~~~~&
\scalebox{1.1}{\includegraphics[height = .08\textwidth, width=.1\textwidth]{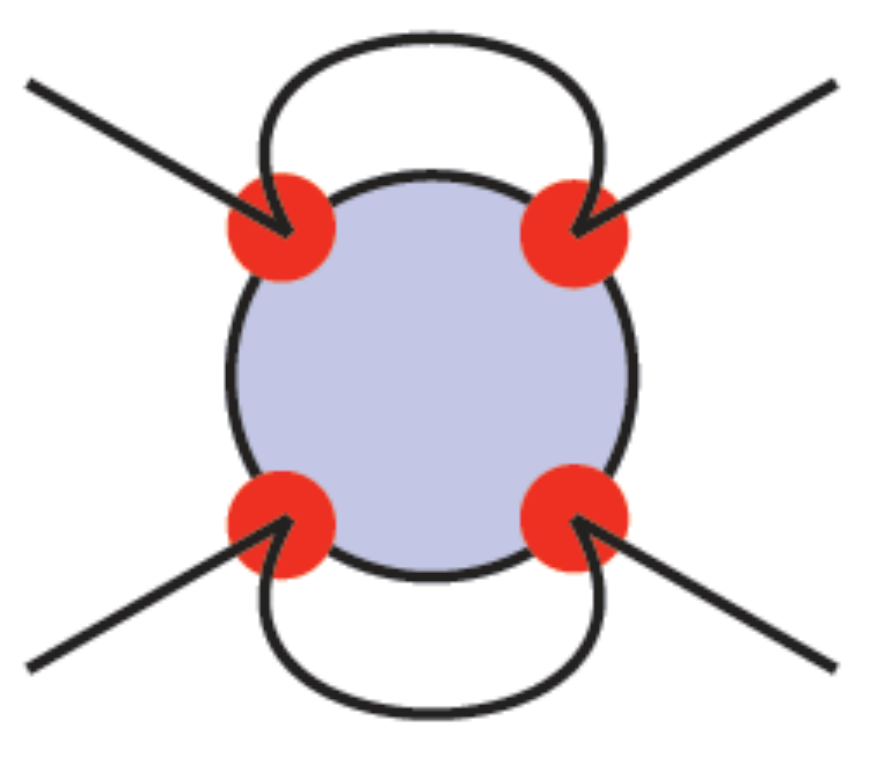}}\\
$0$&$0.02083$&$-0.06893$\\
&&\\
\scalebox{1.1}{\includegraphics[height = .09\textwidth, width=.08\textwidth]{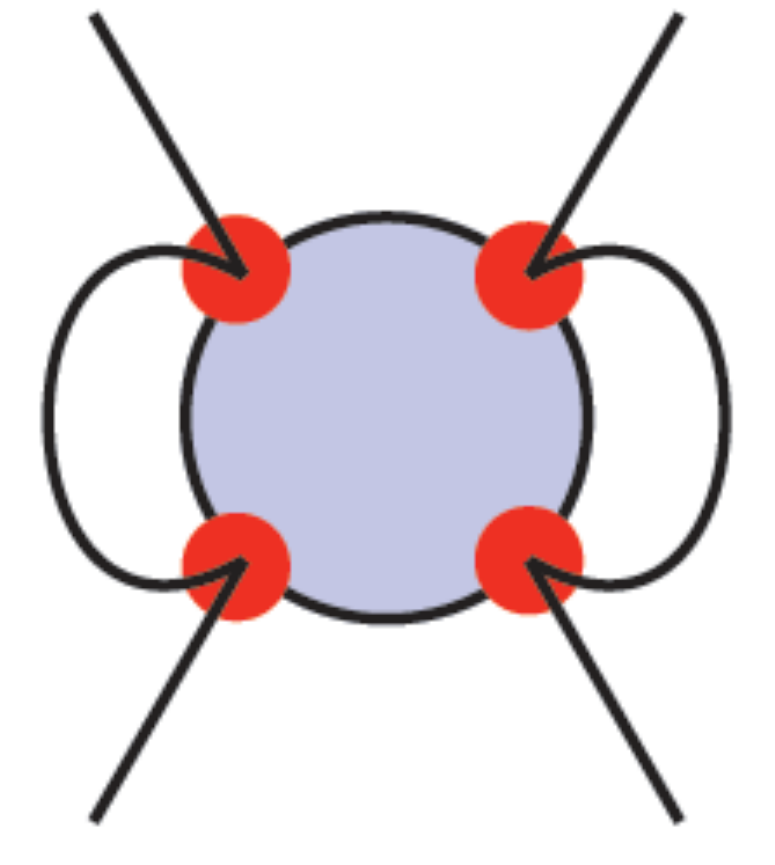}}~~~~~&
\scalebox{1.1}{\includegraphics[height = .09\textwidth, width=.22\textwidth]{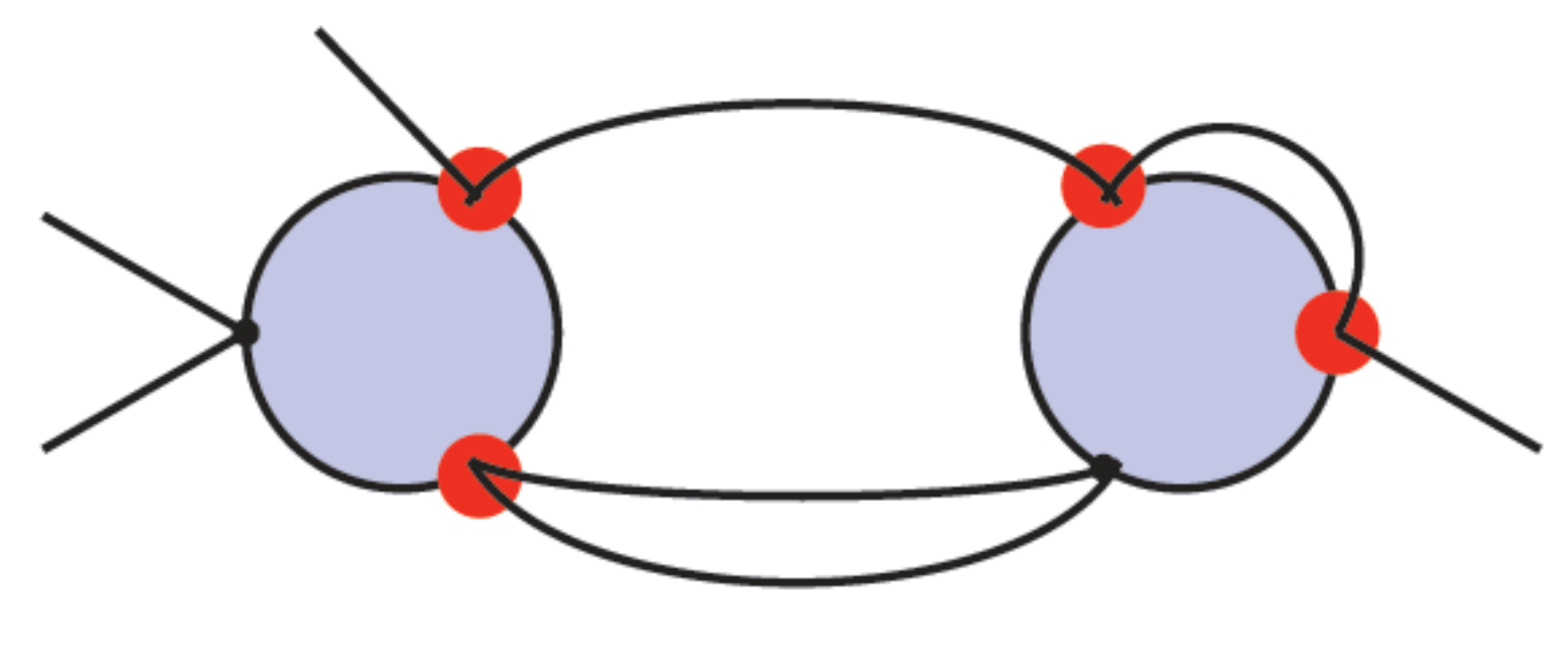}}~~~~~&
\scalebox{1.1}{\includegraphics[height = .085\textwidth, width=.2\textwidth]{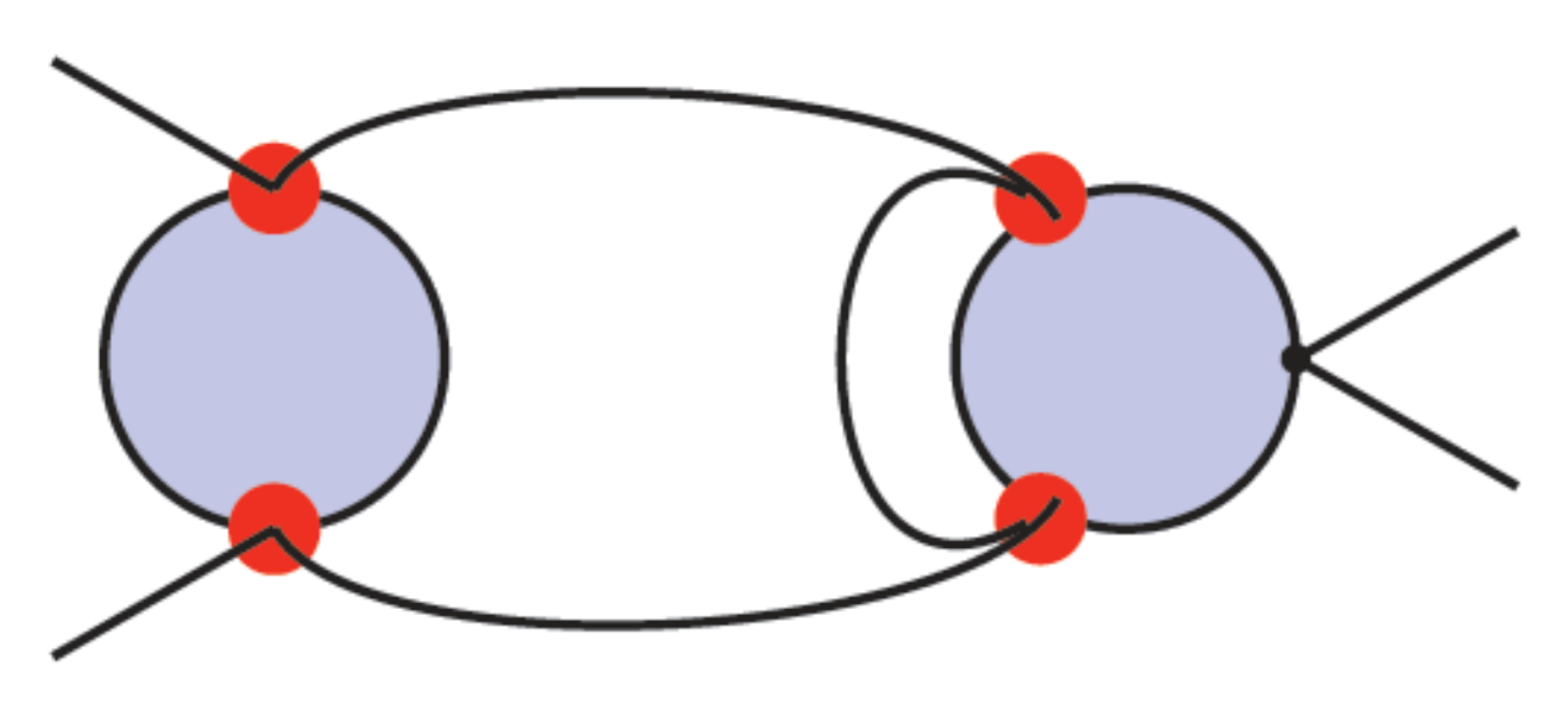}}\\
$-0.01216$&$0.06824$&$ ( - 0.1037)~{\rm to}~(-0.166)$\\
\end{tabular}
\vskip .05in
Figure 5. Corrections to the low momentum limit of the $F^{(2)}$ vertex
\end{center}
\end{table}

Let $C_n$ denote the partial sum of corrections up to terms with
$(m/E_k)^n$, starting with $C_0 = 1.1308$ from the recursive procedure
(\ref{exp4}). We then find
$C_1 = 0.5496$, $C_2 = 0.2730$, $C_3 = 0.0373$. There is a small ambiguity in an integral for the last diagram in figure 5 \cite{KNY}, so that $C_4$ is in the range
$-0.05843$ to $-0.00583$.
The partial sums are systematically decreasing in value, showing that the ordering of diagrams by powers of $m/E_k$ does constitute a sensible expansion.
Also, the corrections to the order we have calculated are small. For the string tension, we find
\beq
\sqrt{\sigma_R} = 
 e^2 \sqrt{ c_A c_R\over
4\pi }~\left\{ 
\begin{matrix}
\bigl( 1 - 0.02799 +\cdots \bigr)\\
\!\!\!\bigl( 1 - 0.0029 +\cdots \bigr)\\
\end{matrix}\right.
 \eeq
This correction, of the order of $-2.8\%$ to $-0.03\%$, is entirely consistent with lattice calculations. Terms of order $(m/E_k)^5$ are expected to contribute at the level of a fraction of $1\%$, 
and likewise for diagrams with two or more current loops.

\section{Discussion}

Perhaps the most important point to be highlighted is the fact that one has a systematic expansion and a scheme for calculations. 
At first glance, the possibility of any expansion scheme in 2+1 dimensional Yang-Mills theory seems very remote, to say the least. The coupling constant $e^2$
does not constitute an expansion parameter. It simply tells us that modes of momenta $\ll e^2$ should be treated nonperturbatively, while modes of momenta $\gg e^2$ can be treated perturbatively; it is at best a marker for this separation. Nevertheless, from what we have discussed, a systematic calculational procedure is possible. This is done by keeping $m$ and $e^2$ as independent until the end, giving a parametric way to classify various contributions to any calculation. Further, there is a kinematic factor, of the form
$(m/E_k)^n$, which helps in the further classification of corrections. The corrections to the lowest order results for the string tension then come out to
be rather small.

Resummation calculations have also indicated some sort of smallness for the corrections. The magnetic mass results quoted came from 
calculations to ``one-loop order" for the Schwinger-Dyson equations. In other words, a particular infinite set of diagrams were summed up which corresponded to
one-loop order viewed as a skeletal diagram. A ``two-loop" extension of this, 
with no obvious reason, gave a reasonably small correction \cite{eberlein}.
Notice also that the values from resummation (coming from the perturbative side) and our value (coming from the large $e^2$ side) are not very different.
There is some sort of stability, almost ``unreasonable stability",  to the calculations in Yang-Mills (2+1).
This suggests that there is a more natural scheme of calculation for this theory,
some features of which
we are discovering, perhaps the hard way.

Thinking in the language of ordinary perturbation theory,
we may note that our analysis involves at least three different resummations. First, the definition of $H$ in terms of $A,~ \bA$ is an infinite series with nonlocal terms. Second, in including $m$ in ${\cal H}_0$,
there is another resummation, since $m$ is really of order $e^2$. Thirdly, we also carry out the summation of contributions due to $F^{(2)}$ to get factors of $m/E_k$.
An {\it a priori} attempt at these resummations, even with some guidance from the Hamiltonian analysis, is clearly very difficult.

Where do we go from here? Higher order corrections to the string tension,
a better computation of glueball masses, extension to the torus and finite-temperature effects as well as the inclusion of matter fields are 
some of the questions we could now explore systematically.

I thank the organizers for inviting me to this very interesting workshop and for accommodating my difficult schedule.

\end{document}